\documentclass[10pt,aps,prb,twocolumn,superscriptaddress,floatfix,showpacs,longbibliography]{revtex4-1}
\usepackage[utf8]{inputenc}
\usepackage{graphicx}
\usepackage{tabularx}
\usepackage[usenames,dvipsnames,table]{xcolor}
\usepackage[version=3]{mhchem}
\usepackage{braket}
\usepackage{upgreek}
\usepackage{hyperref}
\usepackage{amsmath, amssymb}
\usepackage[normalem]{ulem}
\usepackage{soul}

\definecolor{mark}{rgb}{0.85, 0.9, 1}

\hypersetup{hidelinks}
\sethlcolor{mark}

\begin{document}

\title{Benchmarking a boson sampler with Hamming nets}

\author{Ilia A. Iakovlev}
\affiliation{Theoretical Physics and Applied Mathematics Department, Ural Federal University, Ekaterinburg 620002, Russia}
\affiliation{Russian Quantum Center, Skolkovo, Moscow 121205, Russia}
\author{Oleg M. Sotnikov}
\affiliation{Theoretical Physics and Applied Mathematics Department, Ural Federal University, Ekaterinburg 620002, Russia}
\affiliation{Russian Quantum Center, Skolkovo, Moscow 121205, Russia}
\author{Ivan V. Dyakonov}
\affiliation{Quantum Technology Centre and Faculty of Physics, M. V. Lomonosov Moscow State University, Moscow 119991, Russia}
\author{Evgeniy O. Kiktenko}
\affiliation{Russian Quantum Center, Skolkovo, Moscow 121205, Russia}
\author{Aleksey K. Fedorov}
\affiliation{Russian Quantum Center, Skolkovo, Moscow 121205, Russia}
\author{Stanislav S. Straupe}
\affiliation{Russian Quantum Center, Skolkovo, Moscow 121205, Russia}
\affiliation{Quantum Technology Centre and Faculty of Physics, M. V. Lomonosov Moscow State University, Moscow 119991, Russia}
\author{Vladimir V. Mazurenko}
\affiliation{Theoretical Physics and Applied Mathematics Department, Ural Federal University, Ekaterinburg 620002, Russia}
\affiliation{Russian Quantum Center, Skolkovo, Moscow 121205, Russia}

\date{\today}

\begin{abstract}
Analyzing the properties of complex quantum systems is crucial for further development of quantum devices, yet this task is typically challenging and demanding with respect to required amount of measurements.
A special attention to this problem appears within the context of characterizing outcomes of noisy intermediate-scale quantum devices, 
which produce quantum states with specific properties so that it is expected to be hard to simulate such states using classical resources.
In this work, we address the problem of characterization of a boson sampling device, which uses interference of input photons to produce samples of non-trivial probability distributions that at certain condition are hard to obtain classically.
For realistic experimental conditions the problem is to probe multi-photon interference with a limited number of the measurement outcomes without collisions and repetitions. 
By constructing networks on the measurements outcomes, we demonstrate a possibility to discriminate between regimes of indistinguishable and distinguishable bosons by quantifying the structures of the corresponding networks. 
Based on this we propose a machine-learning-based protocol to benchmark a boson sampler with unknown scattering matrix. 
Notably, the protocol works in the most challenging regimes of having a very limited number of bitstrings without collisions and repetitions. 
As we expect, our framework can be directly applied for characterizing boson sampling devices that are currently available in experiments. 
\end{abstract}

\maketitle

\section*{Introduction}
The idea behind quantum computing is to manipulate complex --- entangled, many-body --- quantum states to solve computational problems~\cite{Brassard1998,Ladd2010,Fedorov2022}.
Certain quantum algorithms use a feature of a possibility to efficiently check the correctness of the obtained results, for example, as it takes place for the Shor's factorization algorithm~\cite{Shor1994}.
In a general case, however, the problem of characterization and verification of quantum states that are produced by quantum computing devices is highly non-trivial, yet it is essential to understand whereas the quantum devices work correctly. 
This task becomes even more challenging taking into account the fact that currently developed quantum processors are highly affected by decoherence, so they belong to the class of noisy intermediate-scale quantum (NISQ) devices.
A celebrated example is a seemingly unresolvable problem of sampling the output of a pseudo-random 53-qubit circuit performed by the Google team~\cite{Martinis2019} with the Sycamore processor, which is exponentially more difficult to do with classical computing.
This breakthrough study stimulates development of the methods not only for efficient simulating large-scale quantum wave function on classical devices~\cite{Pednault2019,Huang2020,PanPan2021,Vazirani2022,Liu2022,Harrow2022}, 
but also approaches for distinguishing quantum states delocalized in the Hilbert space from each other with a very limited number of measurements \cite{Sotnikov2022, Heyl2023}.
Recent random circuit sampling experiments \cite{morvan2023phase} with 70 qubits define a new boundary for demonstrating quantum advantage.

In addition to the gate-based model of quantum computing, remarkable progress with developing of boson sampling (BS)~\cite{Aaronson2013,Gogolin2013,Tichy2014} has been performed~\cite{Li20192}
(starting by first experimental realizations~\cite{Tillmann2013,White2013,Walmsley2013}). 
Currently, BS represents a popular quantum playground for testing novel approaches~\cite{Wang2016,Walschaers2016,Huang2017,Chabaud2021,Kuo2022}, 
where one faces a certification problem for a photon device with exponentially large output state space in the absence of a classical counterpart imitated with a classical computer~\cite{Rubtsov2021,Smelyanskiy2021}.
More specifically, for a given device that takes $n$ photons as an input and allocates them over $m$ output modes according to some probability distribution function, 
one should be able to certify that outcome data arise from  indistinguishable photons with limited number of measurements. 
Recent experiments on large-scale boson sampling have been used to demonstrate quantum computational advantage~\cite{Pan2020,Pan2021,Madsen2022, deng2023gaussian}.
The certification of a boson sampler generally assumes, first, unambiguous distinguishing it from a classical device that generates outcomes according to a distribution, for instance uniform one. 
Moreover, a related problem is to define whether given sets of samples were drawn from the same boson sampler or different ones~\cite{Wang2016}. 
Finally, from practical perspective recognizing the regimes of indistinguishable bosons to distinguishable ones, when performing a limited number of experiments with a boson sampler is also of great importance.

\begin{figure*}[!ht]
    \includegraphics[width=\textwidth]{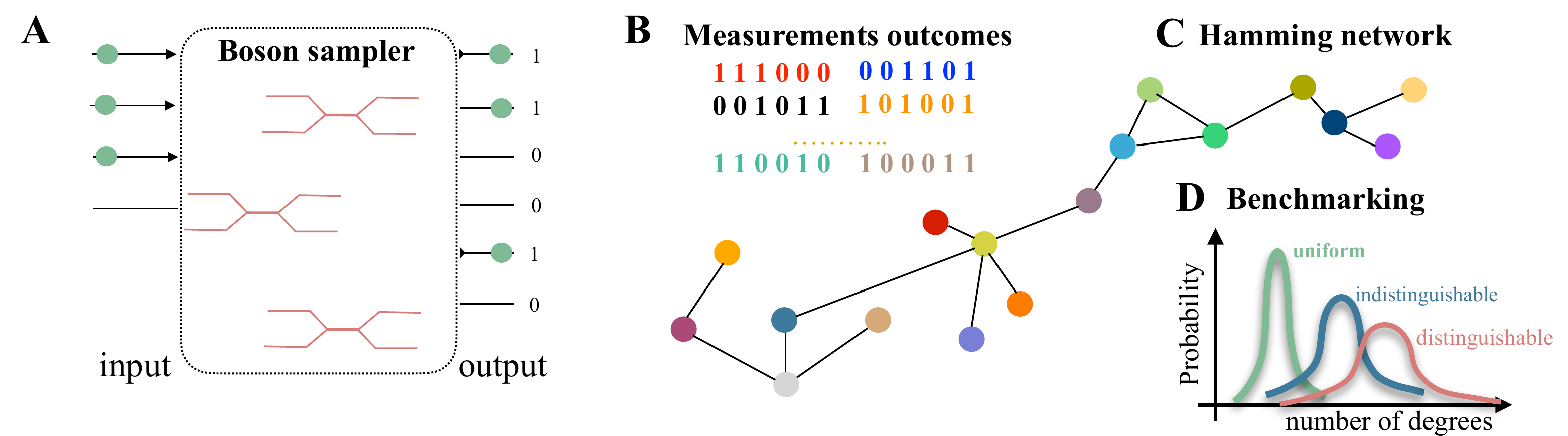}
	\caption{\label{intro} Protocol for constructing a Hamming network and certification of a boson sampler. (A) Schematic representation of the boson sampler that takes 3 photons as input and distribute them over output modes according to some probability distribution function. (B) A limited set of outcomes (bitstrings) without repetitions and collisions as obtained from the boson sampler. (C) Construction of the network in which each node corresponds to the specific bitstring. If the Hamming distance between two bitstrings is smaller or equal than the chosen cutoff radius $R$, then the corresponding nodes are connected. For each node the total number of links (degrees) is calculated. (D) Distribution of the nodes (bitstrings) with respect to the number of degrees allows  discriminating between sets of uniform, distinguishable and indistinguishable boson bitstrings.}
\end{figure*}

There are, generally speaking, two widely used strategies for solving the problem of verification of a boson sampler. 
The first one assumes that one exploits an insider information on the boson sampler in question. 
For instance, it could be details of a scattering matrix $U$ describing the connection between input and output modes~\cite{Spagnolo2014,Pan20192}. 
The existence of a trusted boson sampler that is assumed in some studies~\cite{Fabio2019} can also considerably facilitate the validation of a photonic device with different clustering techniques. 
The second class of approaches for certifying boson samplers fully relies on the analysis of the measurements outcomes. 
For instance, statistical benchmarking proposed in Ref.~\onlinecite{Walschaers2016} is based on the calculation of  pair correlation functions for all possible output modes combinations.
While, theoretically, these correlators allow one to probe many-particle interference, 
the practical realization of such a benchmarking requires some experimental efforts in performing numerous measurements for all possible inputs and should be verified in each case. 

The complexity of the boson sampler certification~\cite{Pan2019,Madsen2022} suggests to make use of the whole arsenal of available methods, including those from completely different fields of research dealing with problem of analyzing complex processes. 
For instance, a fresh look at the problem can be taken with machine learning (ML) techniques including clustering methods \cite{Wang2016,Fabio2019,Giordani2023}, 
combination of the low-dimensional representation and convolutional neural network~\cite{Flamini2019} or others. 
While the main focus in these ML-based studies is on the Hamming distance (or $L_1$ norm) between bitstrings $b_i=(b_i^1,\ldots,b_i^m)$ and $b_j=(b_j^1,\ldots,b_j^m)$ that is defined as $\mathcal{D}_{ij} = \sum_{k=1}^{m} |b^k_i - b^k_j|$, 
as it has been shown, taking into account collisions as well as bitstrings statistics also plays an important role in validating boson sampler. 
Remarkably, discriminating sets of boson sampler outcomes without both collisions of photons and repetitions of events has not been demonstrated up to date, which actually corresponds to a typical experimental situation. 

In this work, we propose and demonstrate a protocol for retrieving meaningful information about photons source, which can be extracted even when collision- and repetition-free sets of boson sampler outcomes are only available.
For this purpose, we use the concept of the Hamming network recently introduced in Ref.~\onlinecite{Heyl2023} for the analysis of the complexity of the quantum wave functions and detecting quantum phase transitions. 
Specifically, we apply it to explore the structure of links in the Hamming network constructed on the basis of the measurements outcomes of a boson sampler (see, Figs.~\ref{intro} A and B). 
Each BS event is associated with a colored node in the Hamming network. 
Since there are no repetitions of the bitstrings, there are no nodes of the same color in Fig.~\ref{intro}C. 
Having chosen a cutoff radius, $R$, that can be in the range between minimal and maximal bitstring distances within the entire ensemble of outcomes, one compares it to the Hamming distances of individual bitstring pairs. 
If $R \geq \mathcal{D}_{ij}$, we connect the $i$th and $j$th events with a link. 
Computing the statistics over the number of links (Fig.~\ref{intro}D) in the network constructed for a given set of measurements outcomes allows characterization of the boson sampler in the situation of information scarcity. 
As it is shown below, one can discriminate between uniform and non-uniform samplers as well as distinguishable and indistinguishable regimes of BS.

\section*{Results}

\begin{figure*}[!ht]
    \includegraphics[width=1.8\columnwidth]{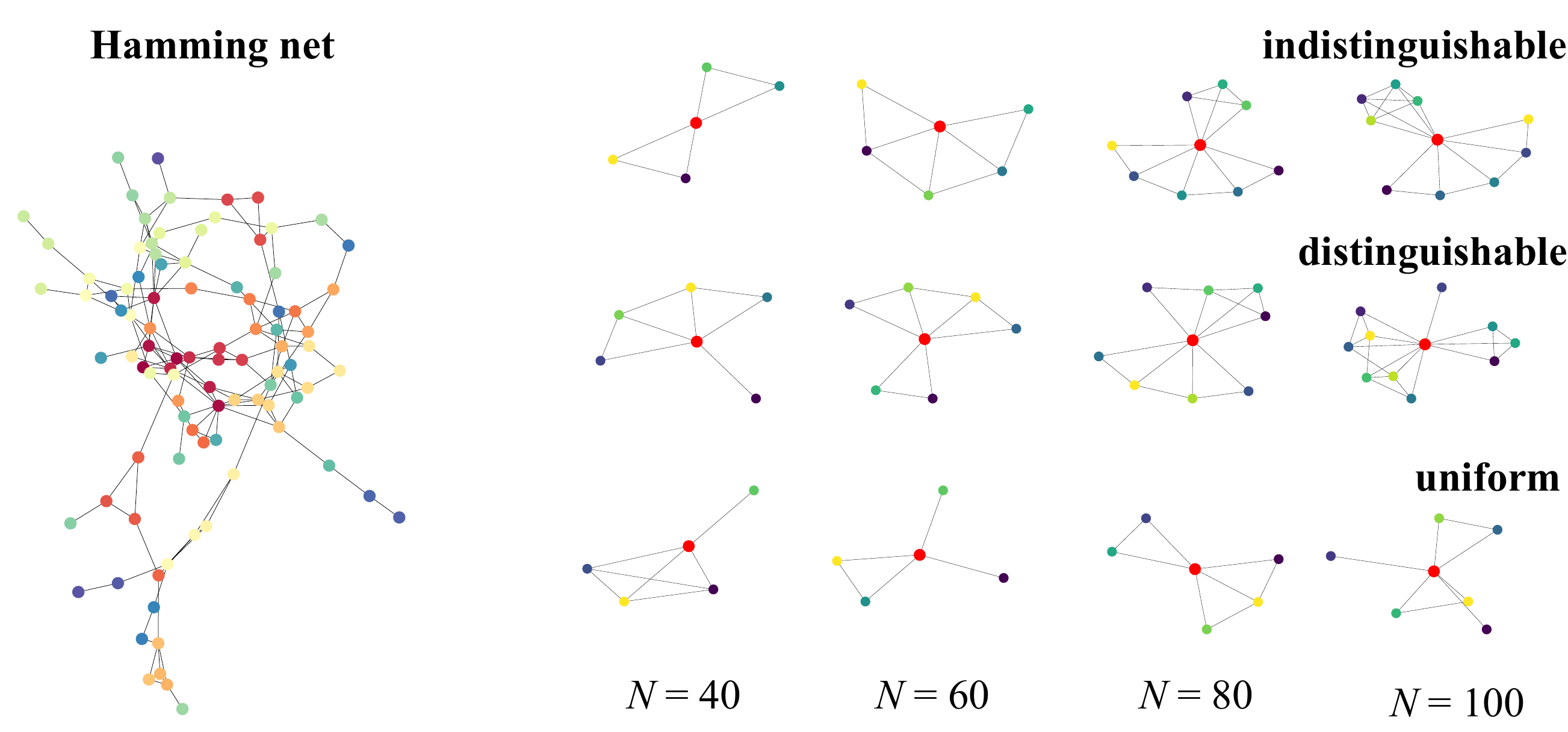}
	\caption{\label{network} (Left) Example of the Hamming network constructed for 100 outcomes taken from the boson sampler with 4 bosons and 16 output modes in the regime of indistinguishable particles. The cutoff radius $R$ is equal to 2. (Right) Fragments of the Hamming networks that contain nodes (red circle) with the largest number of degrees. These networks were constructed with $N =40$, $60$, $80$, and $100$ bitstrings from indistinguishable (top row), distinguishable (middle row) and uniform (bottom row) photons sources.}
\end{figure*}

\subsection*{Constructing Hamming networks}
We start our consideration with constructing Hamming networks for the collision-free sets of bitstrings that were generated with loading $n$ photons into the boson samplers of $m = n^2$ output modes, where $n$ = 4, 5, 6 and 7. According to Ref.\onlinecite{Aaronson2013} it is believed that such a quadratic dependence of modes number on $n$ corresponds to the lower bound for demonstrating quantum advantage with boson sampling. In these BS settings the total number of the unique collision-free outcomes is defined as $C_m^n$ and equal to 1820 ($n=4$), 53130 ($n=5$), 1947792 ($n=6$) and 85900584 ($n=7$). By collision-free set we mean that all $n$ photons are detected in distinct output modes. This regime is consistent with commonly used photo-detectors that do not allow for photon number discrimination. From the point of view of real boson sampling experiments the considered configurations are realistic and imitate characteristics of the state-of-the-art devices~\cite{photonWang2017, PhysRevLett.121.250505, Pan20192}. A detailed technical information concerning the boson sampler simulator we use is given in the Supplementary Information. We would like to stress that each outcome (bitstring) is unique within the particular set, which excludes using outcomes statistics for recognizing many-particle interference regime realized in BS. It makes the approaches relying on the choice of the states with highest probability~\cite{Wang2016, Fabio2019} out of game and strongly motivates to employing methods~\cite{Heyl2023} that can reveal hidden dependencies, structures and correlations in a limited amount of data.   

\begin{figure}[!ht]
    \includegraphics[width=\columnwidth]{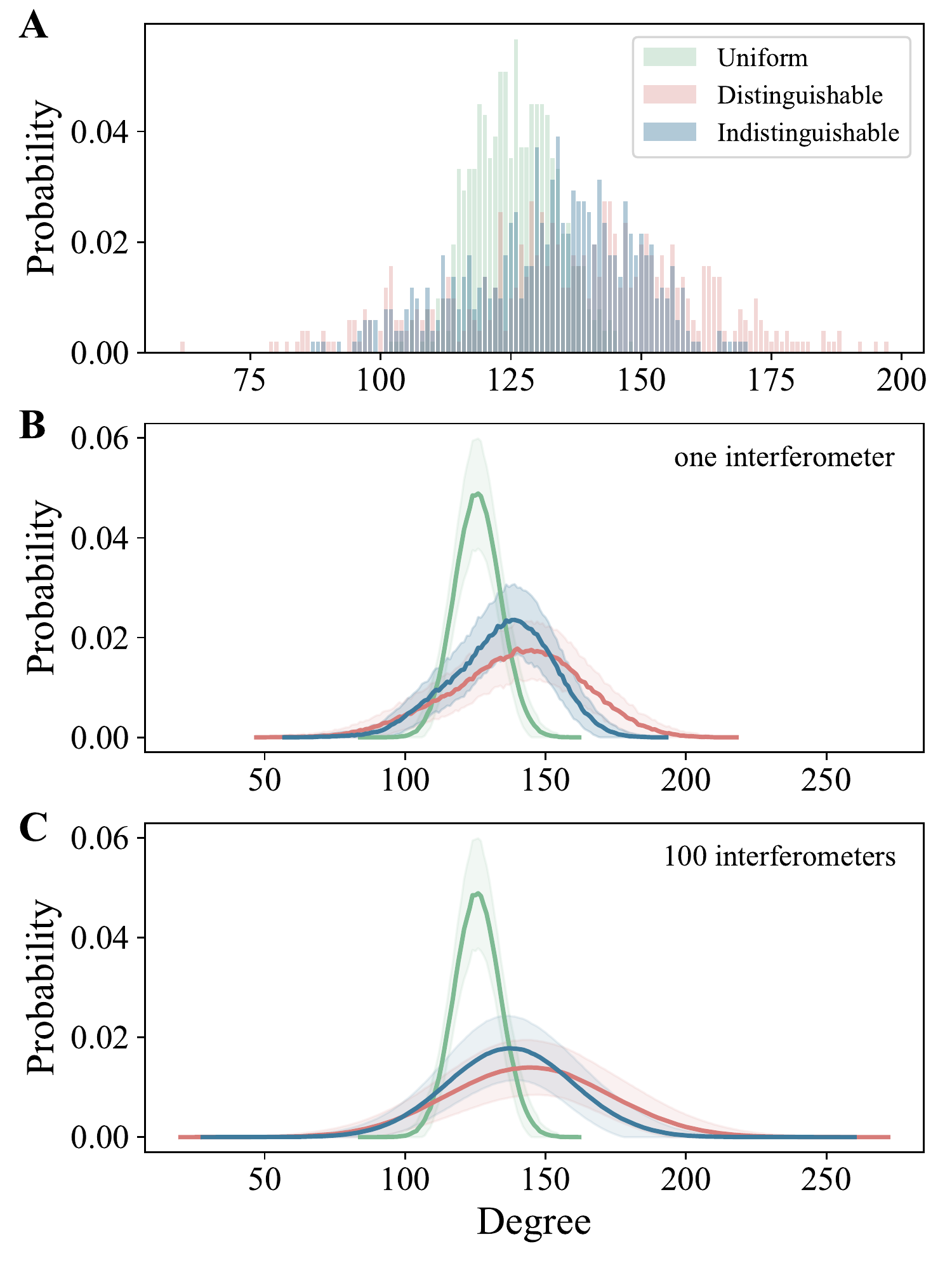}
	\caption{\label{fig:distributions} Comparison of the degree distributions estimated from the Hamming nets that were constructed with uniform (green), distinguishable (red) and indistinguishable (blue) samplers outcomes. These results were obtained for boson sampler with 16 modes and 4 photons and cutoff radius $R$ = 4. (A) Comparison of the distributions on the level of single sample of 512 bitstrings. (B) Data averaged over 512 samples each containing 512 bitstrings obtained from the same interferometer. (C) Results averaged over 100 interferometers. Data for each interferometer was averaged over 512 samples with size $N$=512.}
\end{figure}

The main parameter when constructing the network for a set of outcomes is the cutoff radius, $R$ that defines the particular structure of the Hamming network. Since all the bitstrings we collect have the same number of ''1'' bits the minimal difference in the Hamming distance, $\mathcal{D}_{ij}$ between two arbitrary chosen bitstrings is equal to 2, which is the minimal step for the cutoff radius change. In Fig.~\ref{network} A we give an example of such a net constructed with $R$ =2 for which all the links have the same weight. In other words, within our approach to differentiate the photon sources we will only use the information about number of degrees. One can see that in case of the small $R$ = 2 there are many nodes with a few links, which can be explained by the small number of bitstrings in the sample and a non-uniform distribution of the corresponding bitstrings over state space. The visualization of network's fragments Fig.~\ref{network} B constructed with minimal cutoff radius and small sets of bitstrings reveals a difference in change of the degrees number depending on the size of the bitstrings set for uniform and non-uniform samplers. In the case of the nets constructed with uniform distribution the number of degrees increases slowly than for distinguishable and indistinguishable ones. At the same time, by these fragments we cannot  discriminate the distinguishable and indistinguishable photons sources. In both cases the graphs obtained for the same number of bitstrings look similar.  

Such a structural difference between networks constructed with uniform and non-uniform samples becomes more evident when analyzing the probability distributions $P_k$ of the network nodes with respect to the number of their degrees $k$. One can think about $P_k$ as a probability that a random node in the Hamming net has exactly $k$ neighbours. The results obtained for individual samples of 512 bitstrings and presented in Fig.~\ref{fig:distributions} A  demonstrate different locations of the means of these distributions. It paves the way for benchmarking outcomes from a non-uniform sampler in fully unsupervised manner. More specifically, for a given set of bitstrings obtained from an unknown sampler we are able to generate the same number of bitstrings distributed uniformly, which can be done efficiently with a classical computer. Then, one constructs networks with bitstrings from unknown device and uniform sampler. By comparing the means and standard deviations of the resulting probability distribution functions one can make the conclusion whether the initial set of bitstrings was generated with uniform or non-uniform unknown sampler. Validity of the proposed certification procedure is confirmed by the results that were averaged over 512 samples taken from the same interferometer (Fig. \ref{fig:distributions} B) and those averaged over 100 independent interferometers (Fig. \ref{fig:distributions} C). In all cases the difference in the distribution properties between uniform and non-uniform samplers is robust.

If one gets information about the scattering matrix $U$ of the device to be certified, there are efficient algorithms such as test of Aaronson and Arkhipov \cite{Aaronson2014} that can validate a boson sampler against uniform on the basis of several outcomes. Importantly, the practical implementation of this test does not assume the calculation of any permanent. In the cases when the details of the multi-particle interferometer are unknown, one can make use of a kind of clustering algorithm. For instance, to characterize a boson device, a bubble clustering protocol \cite{Wang2016} proposed by Wang and Duan determines the structure of the bit-string sample by utilizing frequency of generating individual outcomes. Our approach is different, since constructing Hamming nets to certify BS device excludes repetitions in the bitstring sample.   

While recognizing uniform samplers is straightforward with Hamming nets, the certification of distinguishable and indistinguishable BS regimes is found to be a more challenging task. As follows from Figs.~\ref{fig:distributions} B and C the averaging over samples and over scattering matrices leads to a strong dispersion (colored regions) of the probabilities functions, $P^{\rm D}_k$ (for distinguishable particles) and $P^{\rm I}_k$ (for indistinguishable particles). In other words, a sampler can give a distribution of the nodes that will strongly differ from the mean probability profile denoted with line in Fig.~\ref{fig:distributions} C. Although, there are some differences in mean and standard deviation values between the averaged distinguishable and indistinguishable data (Fig.~\ref{fig:distributions} C), such differences are too weak to be used by a researcher for a manual benchmarking of a boson sampler. It motivates us to develop a machine learning protocol for certifying boson samplers as described in the next section. 

\subsection*{Machine learning BS regimes}

The theoretical description of a system or a process with a limited number of observations available for a researcher is a standard task in science that may arise in various fields. In physics, classifying different types of Brownian motion with short trajectories~\cite{Brownian1, Brownian2}, constructing phase diagram on the basis of limited number of system's snapshots~\cite{Melko,profiles}, approximating ground state of a quantum Hamiltonian on a quantum computer \cite{Sotnikov_2020} and certifying a quantum state on a quantum device by means of a few measurements~\cite{Sotnikov2022} represent only a few notable examples of problems among many others. Remarkably, in these and other cases machine learning (ML) has turned out to be a very valuable alternative to standard techniques and advance the corresponding fields of research. In this sense, the boson sampling is no exception and there are various machine learning based schemes for benchmarking boson devices. They include basic clustering ML algorithms~\cite{Duan2019} and neural network approaches~\cite{Flamini2019} as well. 

\begin{figure}[!ht]
    \includegraphics[width=\columnwidth]{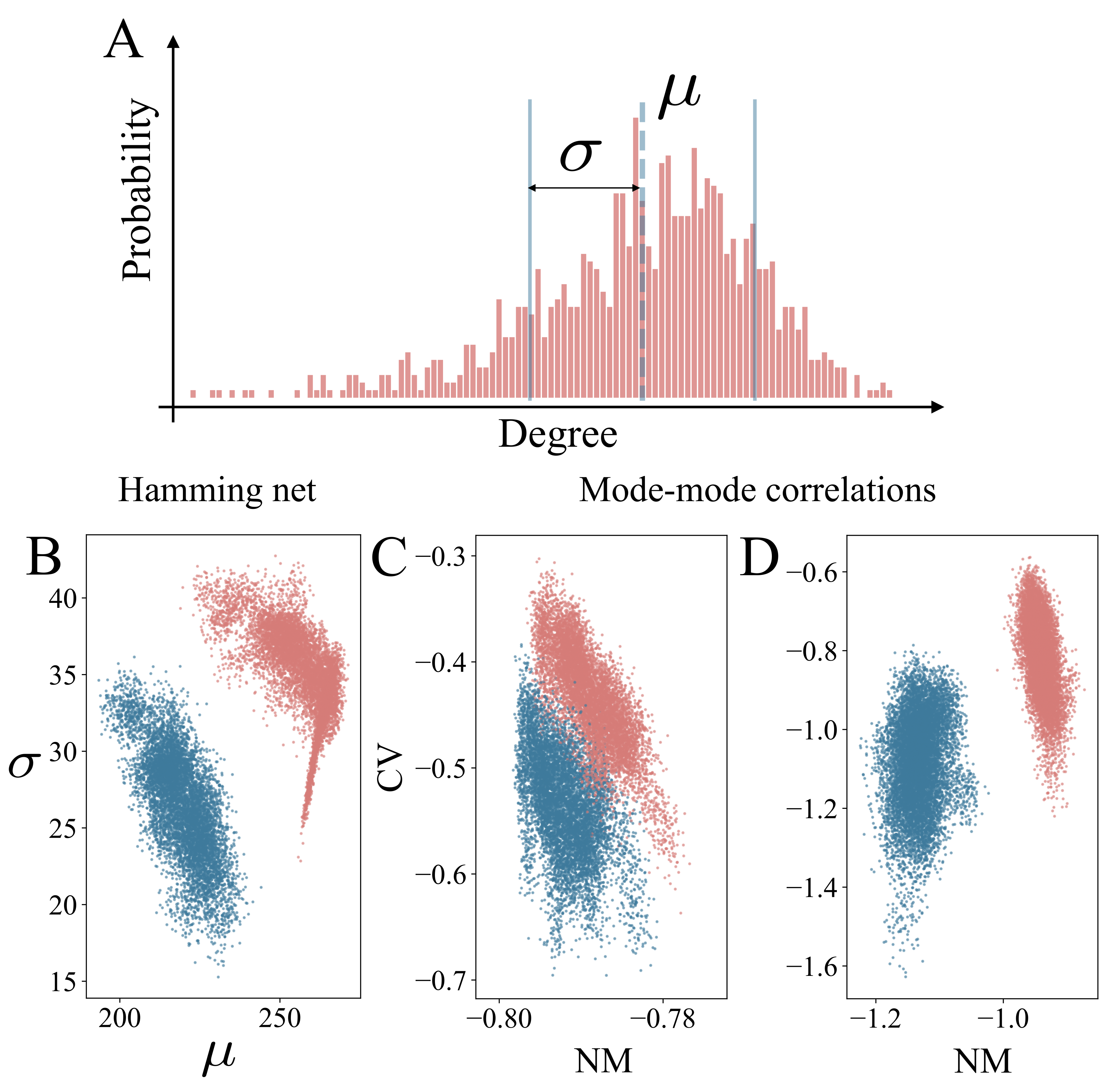}
	\caption{\label{fig:clouds} (A) Schematic representation of the $P_{k}$ distribution obtained using $N = 1024$ bitstrings at $R=6$. (B-D) Low-dimensional visualisations of the configurations taken from the training sets for $n=4$ distinguishable (red) and indistinguishable (blue) bosons and $m=16$ output modes, $N=1024$ and the cutoff radius of 4. Features were taken from the Hamming network (B) and correlation functions (C)-(D) approaches. The last data set (D) includes collisions and repetitions. NM and CV denote the normalized mean and coefficient of variation which are the features introduced in Ref.\onlinecite{Walschaers2016} within mode-mode correlation function approach.}
\end{figure}

In our case discriminating devices with distinguishable and indistinguishable photons on the level of the Hamming networks can be also advanced with basic machine learning. In order to show that, we first perform a feature selection procedure. We find out that the most reliable features are two first moments of the $P_{k}$ distributions for such intermediate cutoff radii $R$ for which $P_{k}$ is Gaussian-like and has both left and right sides (Fig.~\ref{fig:clouds} A). Namely, we use $R=\{2, 4, 6\}$, $R=\{2, 4, 6, 8\}$, $R=\{4, 6, 8, 10\}$ and $R=\{6, 8, 10, 12\}$ for the samplers with $n$ = 4, 5, 6 and 7 photons, respectively. More details on the radii choice are given the Supplementary Information. By the example of the standard deviation and mean features presented in Fig.~\ref{fig:clouds} B one can see that the clouds formed from distinguishable and indistinguishable outcomes are well separated for $n=4$, which, as we will show below, provides a high accuracy in classification with ML. 

At this stage, it is important to recall other quantities that are based on a mode-mode correlation function used in the previous theoretical \cite{Walschaers2016, Huang2017, Flamini_2020} and experimental \cite{Madsen2022} works for benchmarking indistinguishable sampler against distinguishable one. Such correlation functions are defined as $C_{ij} = \langle n_i n_j\rangle - \langle n_i\rangle \langle n_j\rangle$, where $n_i$ in the photon number in the $i$th mode. The features (normalized mean, coefficient of variation, skewness) extracted from distribution of the calculated $C_{ij}$ allow discriminating different samplers. As it follows from Figs.~\ref{fig:clouds} C our consideration challenges the previous results, since the clouds formed by different samplers overlap in the feature space for bit-string sets subjected to additional selection. Comparing Figs.~\ref{fig:clouds} C and D clearly shows that outcomes repetitions and collision events play a crucial role in forming well-separated clouds on the level of features. Thus, the regime without collisions and repetitions for $n=4$ we explore in this work is of particular difficulty. 

\begin{figure}[!hb]
    \includegraphics[width=\columnwidth]{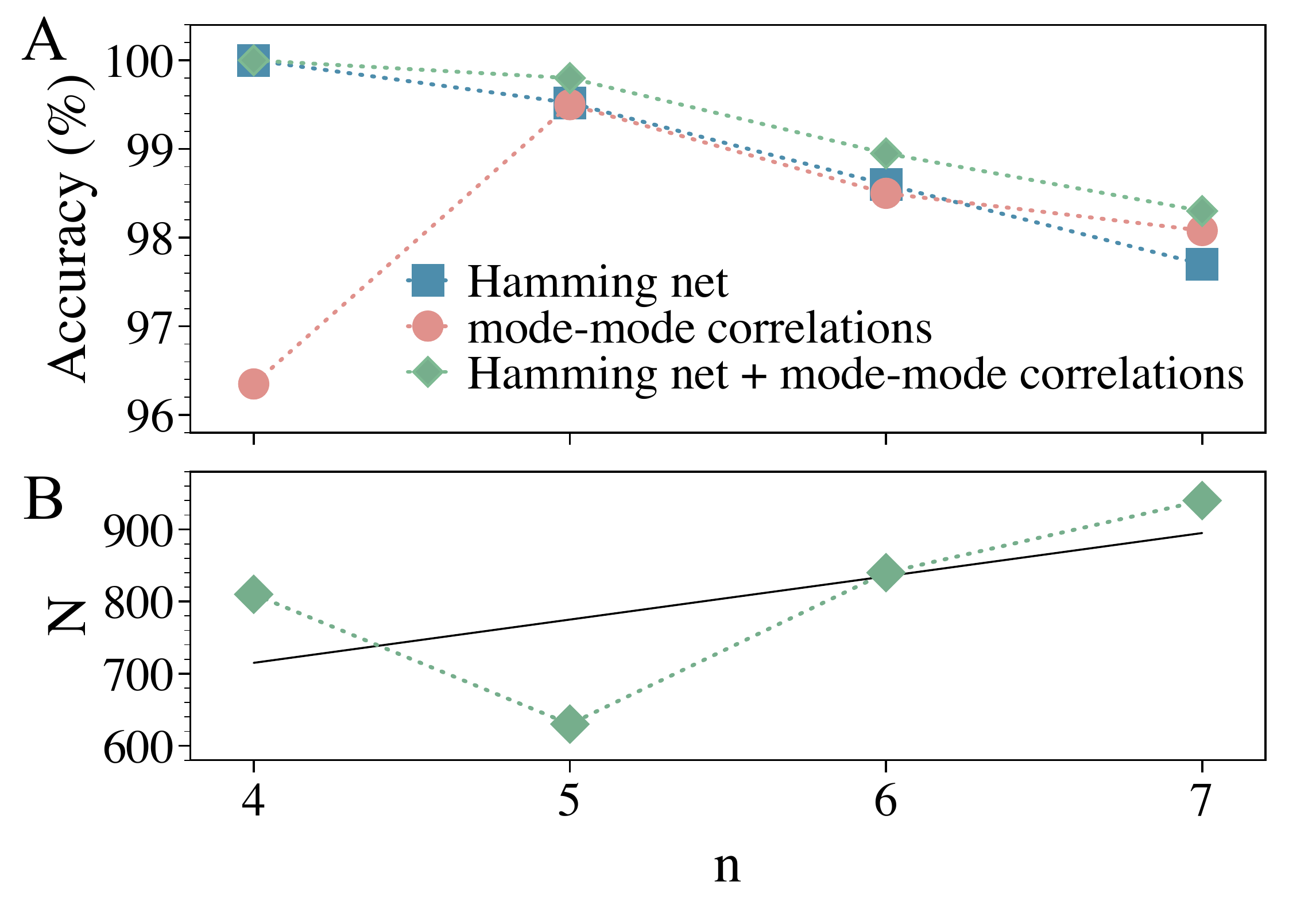}
	\caption{\label{fig:ML} Machine learning certification of boson samplers. (A) Accuracy of the logistic regression model on the testing data obtained from the unseen $U$ matrices. The results are obtained using $N=1024$ bitstrings in each sample. The error bars are smaller than the symbol size. (B) Dependence of the number of unique bitstrings $N$ required to reach the accuracy $\sim 98\%$ on $n$. Features were taken from both Hamming network and mode-mode correlations. The amount of output modes is equal to $m=n^2$. }
\end{figure}

Since separating clouds in the feature plane is a kind of trade-off in our choice of the bit-string number in the sample with respect to the total size of the state space, it becomes important to implement the machine learning to control classification quality for boson samplers with $n > 4$. Among the basic ML methods we tested (see Supplementary Information for more details) the best benchmarking results on scattering matrices unseen during the training stage were achieved with logistic regression (LR). From Fig.~\ref{fig:ML} A one can see that the accuracy of the classification based on the Hamming net features gradually degrades as the number of bosons increases. For instance, in the case of $n=7$ it is about 97.7\% (Fig.~\ref{fig:ML} A), which clearly indicates the smallness of the sample size ($N$ = 1024) with respect to the total state space dimension (85900584). 

In turn, the ML models based on mode-mode correlations demonstrate a considerable enhancement of the classification accuracy for $n=5$ in comparison with $n=4$. It can be explained by a larger effective separation of the centers of the clouds in low-dimensional representation of the data for distinguishable and indistinguishable particles. Nevertheless, the fact that there is the deviation from the ideal 100\% certification of the particle type evidences non-zero overlap of the corresponding clouds. Fig.\ref{fig:ML} A shows ML accuracy for mode-mode correlators that behaves similarly to that obtained with Hamming net data for $n>4$. It has to be stressed out one more time that, the previous works based on the calculation of the mode-mode correlation functions did not develop an intuition about the performance of this approach in the case of the collision-free and repetition-free regime and for the BS setting with $m = n^2$ dependence, which is one of the goals of this work. 

Importantly, both the Hamming nets features and mode-mode correlation features are not correlated by the construction. This fact means that these features can be combined to achieve the better performance in distinguishing boson samplers with machine learning. Indeed, in this case the resulting ML accuracy increases to $\sim 98.3\%$ for $n=7$. These ML results clearly show that accurate benchmarking of unknown boson samplers is possible with a very limited number of bitstrings without collisions and repetitions. 

Another important question is how the required amount of unique bitstrings scales with $n$ if we fix the accuracy. As can be seen from Fig.~\ref{fig:ML} B $N$ grows rather linearly and the obtained values are experimentally reachable which indicates the viability of the proposed approach. Here we fix the accuracy to be $\sim 98\%$. The reason why the BS with $n=5$ requires less bitstrings to reach the same accuracy can be explained in the following way. On the one hand, it has more output modes and therefore more degrees of freedom than the BS with $n=4$, which increase the difference in statistics of the considered regimes. On the other hand, the used $N$ covers the larger portion of the whole basis than in the case of $n\ge6$.

\subsection*{Three distinct measures of BS complexity}
Characterizing photon interferometry is closely related to the problem of describing the sampling complexity. Traditionally, the main focus is on an exponential separation between quantum and classical sampling times, which is considered to be one of important examples for demonstrating the quantum advantage. However, this is only one of the possible measures of how hard it is to create a sample with a given boson device (computational complexity). A detailed comparison of the distinguishable and indistinguishable bitstrings sources can enrich our understanding, not only in terms of sampling complexity but also in relation to the structure of the generated data, the quantification of information content, and others. In fact, there are more than 42 different measures \cite{Lloyd,complexity} of the complexity that can be potentially used to characterize a system.

\begin{figure}[!ht]
    \includegraphics[width=0.99\columnwidth]{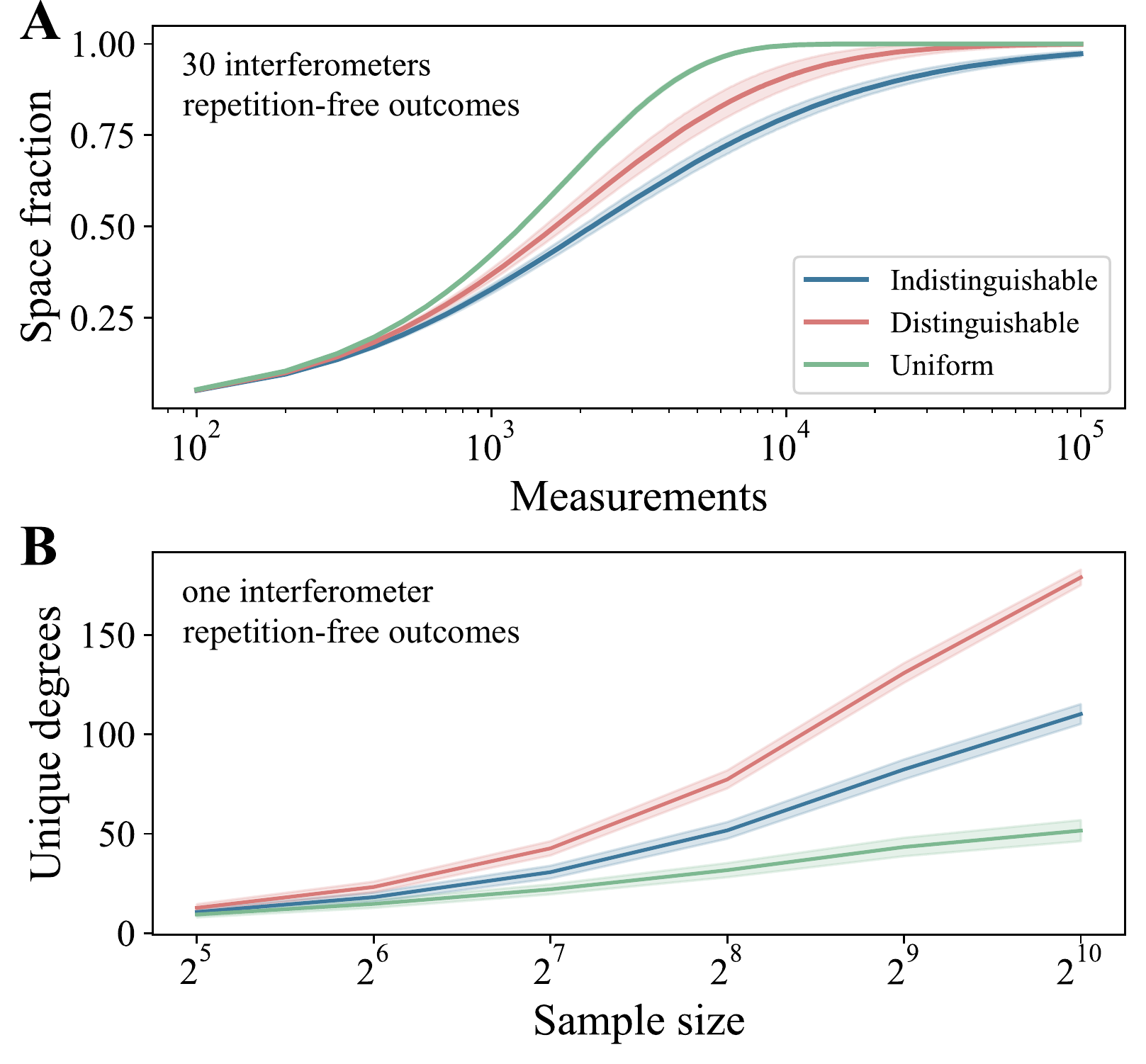}
	\caption{\label{complexity} Boson sampler complexity measures we analyze in this work. (A) Fraction of the states from the Hilbert space as function of performed measurements for uniform, and boson sampler with 16 modes and 4 photons in regimes of distinguishable and indistinguishable photons. Curves for boson sampler were averaged over 30 scattering matrices. (B) Unique degrees count of the sampler network graphs for distinct cutoff radius $ R = 4 $.}
\end{figure}

In this work, we propose three different measures to discriminate outcomes of the boson samplers with respect to the time computational complexity, complexity of the data structure and complexity of describing the information content. The first one can be analyzed by the example of Fig.~\ref{complexity} A that shows the difference between distinguishable and indistinguishable samplers in fraction of the state space that is captured when performing a different number of measurements. One can see that it is more difficult to collect unique bitstrings in the case of an indistinguishable sampler than a distinguishable one with the same number of measurements. The largest difference is observed for $10^4$ measurements. It can be understood from the fact that the probability distribution of the distinguishable bitstrings is closer to uniform than indistinguishable one.

The second complexity measure that provides a robust quantitative characterization and discrimination of the constructed networks with respect to their structure is the number of the unique degrees of the network. In Fig.~\ref{complexity} B we compare the dependencies of the number of the unique degrees on the sample size for distinguishable, uniform and indistinguishable particles. These results evidence that for a given scattering matrix a robust discrimination of the multi-photon regimes can be fulfilled at the cutoff radius $R$ = 4 for the samples that are characterized by minimal sizes of 2$^7$, which is smaller than the total size of state space of 1820.  Importantly, the dispersion of the calculated dependencies averaged over different samples is almost insensitive to the number of the bitstrings in the sample. However, as it was shown above the averaging over different $U$ matrices leads to a strong overlap of the unique degrees calculated with distinguishable and indistinguishable outcomes, which prevents us from using such a measure for certifying BS in an unsupervised manner.   

\begin{figure}[!t]
    \includegraphics[width=0.99\columnwidth]{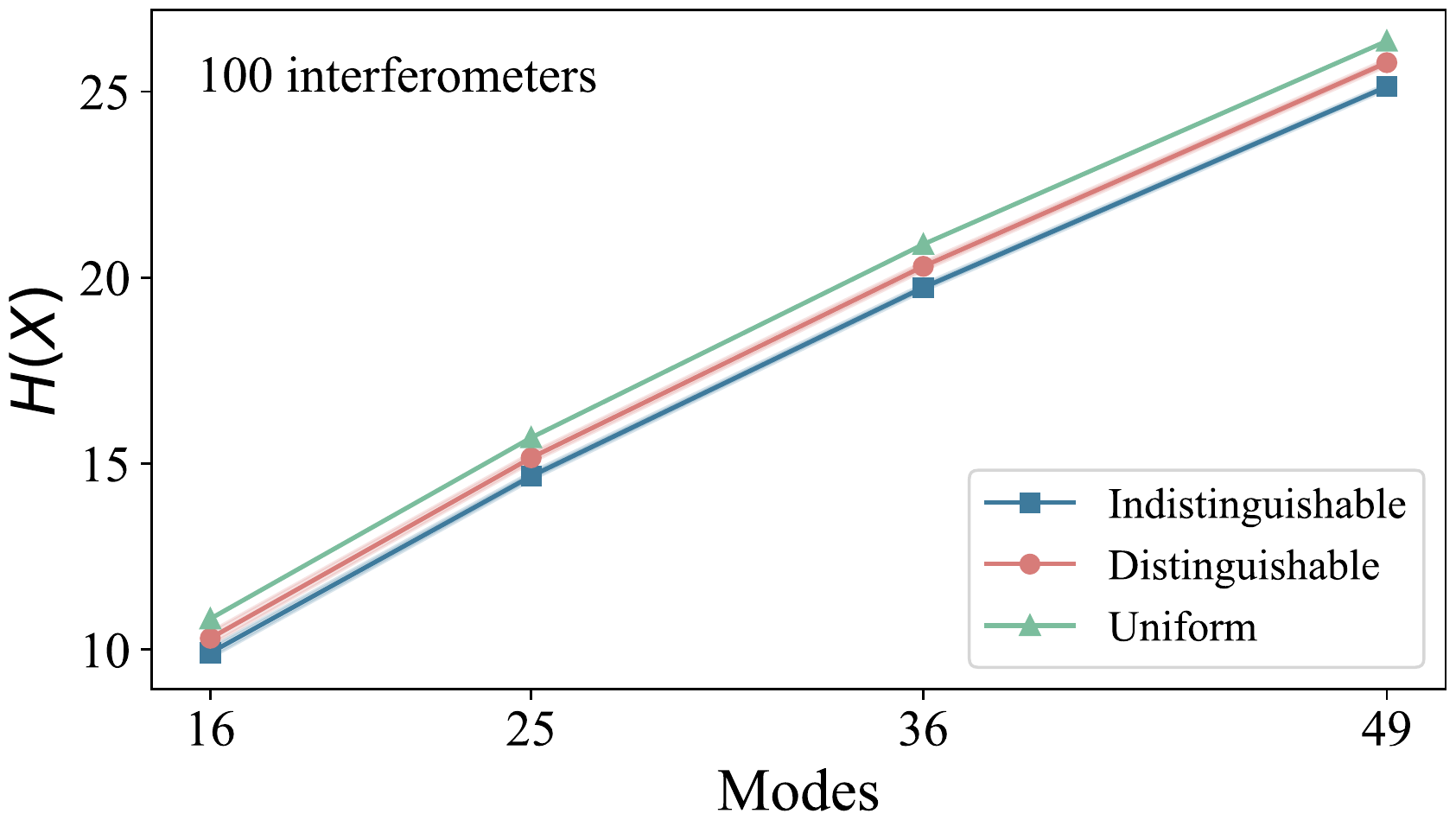}
	\caption{\label{complexity_shannon_entropy} Shannon entropy of probability distributions calculated for boson samplers with distinguishable and indistinguishable photons. The green line denotes the case of the uniform sampler. The data were averaged over 100 boson samplers with different scattering matrices.}
\end{figure}

The third measure to characterize BS complexity we develop in this work is aimed to quantify information content produced by a boson sampler. It requires the account of the outcomes probabilities, which means that {\it we go beyond the consideration above and remove the restrictions on the total number of measurements and the repetitions of the bitstrings}. However, the measurement outcomes are still considered to be collision free. Naturally, the first candidate to describe the information aspect of the BS complexity is the Shannon entropy, $H(X) = - \sum_x p_x \log_2 p_x$, where $p_x$ is the probability of generating particular bitstring $x$ (here $X$ denotes a random variable of obtained bitstring). The Shannon entropy estimates the optimal compression of data~\cite{NielsenChuang2000} that may be achieved for a given source. The obtained results (Fig.~\ref{complexity_shannon_entropy}) show that the Shannon entropy is scaled linearly with respect to the number of output modes. For each BS setting the value of $H(X)$ calculated with the uniform sampler probabilities is nothing but the logarithm of the corresponding state space size.  Unfortunately, a weak difference between the Shannon entropies calculated with probability distributions of distinguishable and indistinguishable photons motivates us to look for another informational measure. 

In this situation, we propose first imitating the BS outcomes with a quantum state that can be initialized on a quantum computer or a quantum simulator. Upon measurements, such a state should reproduce the bit-string statistics of the collision-free boson sampler. It allows one to use the whole arsenal of quantum information theory measures to characterize such a wave function and, as we will show below, to quantify the difference between distinguishable and indistinguishable photons sources. 

At the level of outcomes a collision-free boson sampler can be imitated by using a system of quantum bits whose state in $\sigma^z$ basis is characterized by a special structure of the basis wave functions. More specifically, the number of ''1'' in each basis vector with non-zero probability is fixed to the number of bosons $n$, while the total number of qubits equals to number of output modes, $m$. In analogy to the notable Dicke states~\cite{PhysRev.93.99}, one can write a random counterpart of such a quantum state as  
\begin{eqnarray}
|\Psi_{n} \rangle = \sum_{j} \alpha_j \mathcal{P}_{j}(|0\rangle^{\otimes m-n} \otimes |1\rangle^{\otimes n}),
\label{eq:Dicke_wf}
\end{eqnarray}
where the sum goes over all possible permutations, $\mathcal{P}_j$ of qubits, $\alpha_j$ is the amplitude of the $j$th basis function.

\begin{figure}[!b]
    \includegraphics[width=0.99\columnwidth]{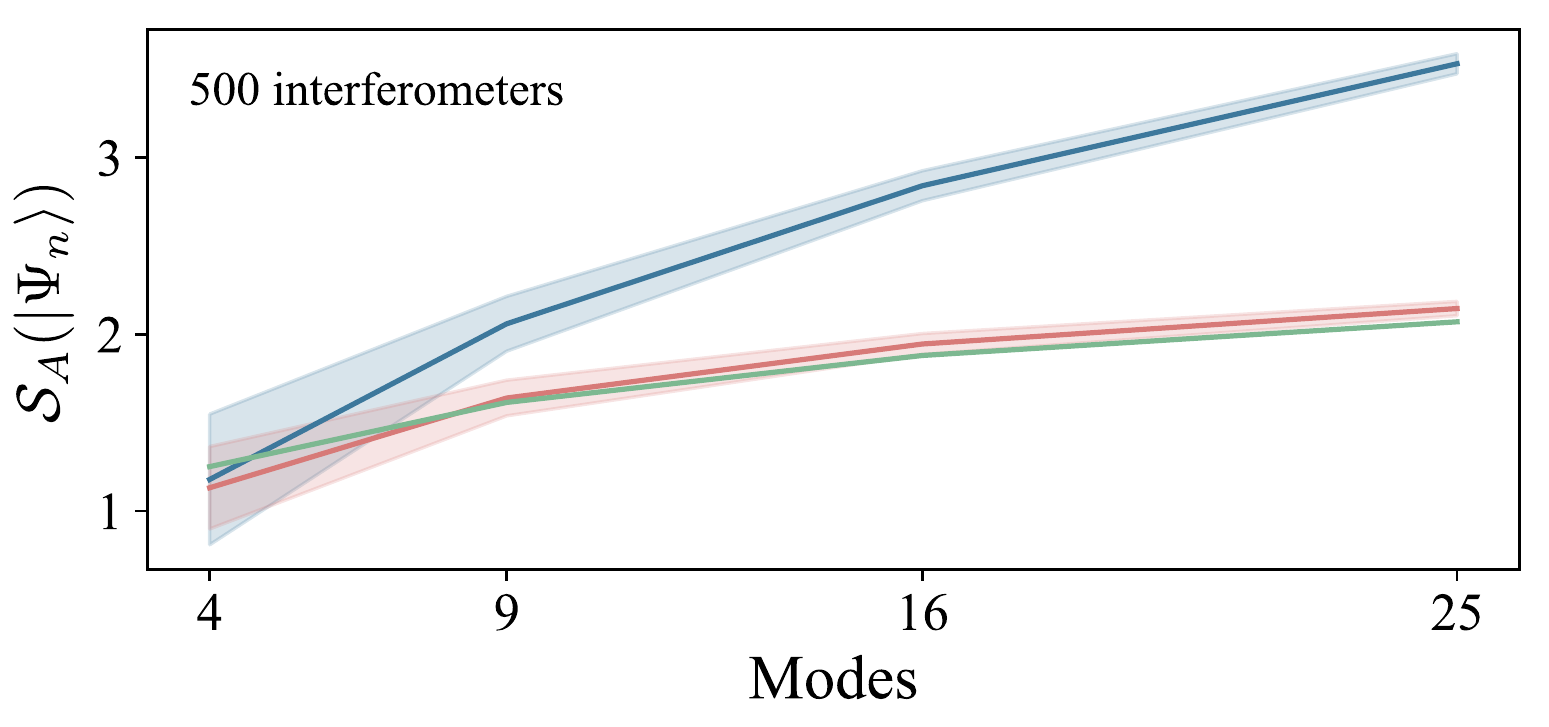}
	\caption{\label{complexity_entropy} Calculated von Neumann entropy for the quantum states that upon projective measurements imitate collision-free outcomes of boson samplers with distinguishable (red line) and indistinguishable (blue line) photons. The data were averaged over 500 boson samplers with different scattering matrices. The green line denotes the uniform sampler results.}
\end{figure}

We would like to stress that $\Psi_{n}$ is not the actual wave function of the boson sampler that should contain information about the complex scattering matrices \cite{Walschaers2016} describing the internal structure of the device. In our case, we aim to reproduce only the BS outcomes having collision-free statistics with such a quantum state. It means that we have an infinite number of choices when defining the coefficients $\alpha_j$ in the wave function Eq.\ref{eq:Dicke_wf} that, in general case, is a complex number with the only constrain $|\alpha_j|^2 = p_j$, where $p_j$ is the probability to generate $j$th bitstring with BS. To simplify the consideration we take $\alpha_j = \sqrt{p_j}$ to be real-valued coefficient. As for the particular complexity measure we choose von Neumann entropy $\mathcal{S}_{A} (\ket{\Psi_{n}}) = -{\rm Tr} [\rho_A \log_2 \rho_A$], where $\rho_A = {\rm Tr}_B \ket{\Psi_{n}}\bra{\Psi_{n}}$ is reduced density matrix for a half-system bipartition into regions $A$ and $B$.

In Fig.~\ref{complexity_entropy} we compare the von Neumann entropies calculated in the case of wave functions imitating outcomes of boson samplers with uniform, distinguishable and indistinguishable particles. In the case of $m \geq 9 $ the averaged data that corresponds to the measure $\mathcal{S}_{A}$ for indistinguishable particles is well-separated from others, which allows one to discriminate this source. The wave functions $\Psi_{n}$ corresponding to the distinguishable and uniform device outcomes are featured with a saturation of the entropy value of around two bits. At the same time the quantum state that reproduces indistinguishable outcomes demonstrates a permanent growth as the number of modes increases. Remarkably, averaging of these results over 500 boson samplers with distinct scattering matrices is characterized by the standard deviation that decreases with increasing the modes number, which paves another way to the accurate classification of unknown boson devices. 

\begin{figure}[!t]
    \includegraphics[width=0.99\columnwidth]{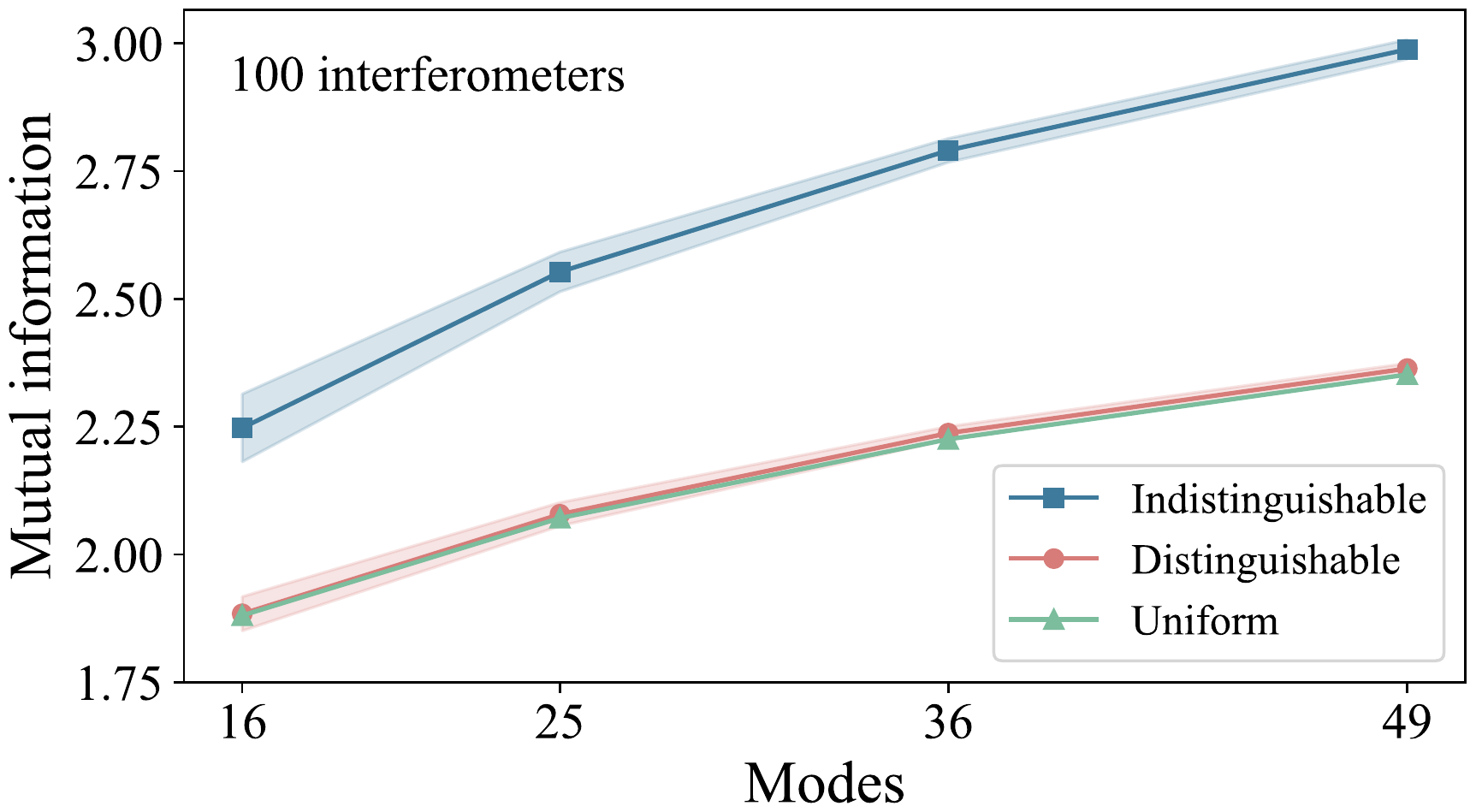}
	\caption{\label{mutual_information} Classical mutual information between two groups of output modes $1,\ldots,\lceil m/2\rfloor$ and $\lceil m/2\rfloor+1,\ldots,m$  
    for boson samplers with distinguishable and indistinguishable photons. The green line denotes the case of uniform sampler. The data were averaged over 100 boson samplers with different scattering matrices.}
\end{figure}

In general case, constructing the quantum state that imitates BS outcomes assumes accumulation of the considerable amount of bitstrings, whose number should be enough to restore probabilities of the basis functions. Then one needs to initialize the quantum state with the particular amplitudes. Thus, the practical realization of the third complexity measure is related to quantum state tomography problem, which is characterized by considerable limitations on the number of qubits in the system in question. However, as it was shown in Ref.~\onlinecite{Torlai2018} implementation of the neural network quantum states can facilitate the solution of the tomography problem for some classes of wave functions, which suggests a distinct way for constructing such entanglement-based BS testers.  

One can also note that the von Neumann entropy ${\cal S}_A(\ket{\Psi_n})$ actually provides a half of the value of the quantum mutual information 
\begin{equation}
\begin{split}
    {\cal}I(A:B)&={\cal S}_A(\ket{\Psi_n})+{\cal S}_B(\ket{\Psi_n})-{\cal S}(\ket{\Psi_n}) \\ 
    &=2{\cal S}_A(\ket{\Psi_n})
\end{split}
\end{equation}
between modes groups $A$ and $B$.
Here ${\cal S}_B(\ket{\Psi_n})$ and ${\cal S}(\ket{\Psi_n})=0$ are von Neumann entropies of modes group $B$ and whole pure state $\ket{\Psi_n}$, correspondingly. 
In Fig.~\ref{mutual_information} we illustrate the behaviour of the corresponding classical mutual information 
\begin{equation}
    J(A:B)=H(A)+H(B)-H(X),
\end{equation}
where $H(A)$ and $H(B)$ are Shannon entropies of bitstrings output at modes groups $A$ and $B$, correspondingly. 
One can see that the values of the classical mutual information drastically differ for indistinguishable and distinguishable photons. 
At the same time, the classical mutual information for distinguishable photons is almost the same for the uniform distribution. Thus, the behaviour of the classical mutual information agrees qualitatively with that calculated on the basis von Neumann entropy in the quantum case.


\section*{Conclusion and outlook}
In this paper, we have developed the procedure for benchmarking boson samplers and distinguishing different regimes of the multi-photon interference using network theory. Our approach is based on constructing networks for limited number of collision- and repetition-free BS outcomes by using the Hamming distances as a parameter that controls the net structure. Already at this level it becomes possible to distinguish uniform and non-uniform BS devices by comparing the degree distributions of the constructed networks. In turn, the certification of the non-uniform samplers into distinguishable or indistinguishable classes is shown to be a more delicate problem that can be solved with machine learning. The performed ML calculations reveal a high accuracy ($>$ 98\%) in classifying distinguishable and indistinguishable photons sources. Importantly, the number of the bitstrings required for such an accurate classification grows linearly as the number of output modes increases while there is an exponential growth of the state space size. 

The proposed scheme for benchmarking boson samplers is mainly based on the information about unique degrees of the constructed networks. At the same time, network links can have different weights depending on the particular values of the Hamming distance between outcomes. Taking this into account would enrich and extend characterization of boson samplers by using network theory. On the other hand, benchmarking with machine learning could also benefit from considering the weights, since it could produce distinct useful features.  

Ignoring the collision events in the BS data allows us to explore a connection between BS sampling and quantum computing. Namely, we derive and characterize random Dicke wave functions that reproduce the statistics over BS outcomes for distinguishable and indistinguishable sources. These quantum states reveal different behaviour in the entanglement depending on the system size. Thus, a distinct research line can be initiated to explore ways including approximation of the wave function with neural network for efficient reconstructing random Dicke states on a quantum device.

From the perspective of the real BS experiments, exploration of the intermediate regimes between limits of indistinguishable and distinguishable particles that are purely theoretical ones is of a particular interest for further development of the Hamming network approach. In this regard, one can combine the proposed machine learning scheme with numerical methods for simulating the photon of a partial distinguishability \cite{Renema2018, Tichy2015}. Thus, the ML models trained in this work can be examined on the samples with predefined degrees of distiguishability. On the other hand, such degrees can themselves be used as labels in machine learning, which assumes the classification with respect to more than two classes as was done in this work. All these steps will facilitate implementation of our approach in diagnosing real BS devices.

\section*{Acknowledgments}
This work was supported by the Russian Roadmap on Quantum Computing (Contract No. 868-1.3-15/15-2021, October 5, 2021). The work of AKF is also supported by the RSF Grant 19-71-10092 (analysis of certain aspects of machine learning applications).

\section*{Supplementary Information}
\subsection*{Boson sampling}

\begin{figure}[!ht]
    \includegraphics[width=\columnwidth]{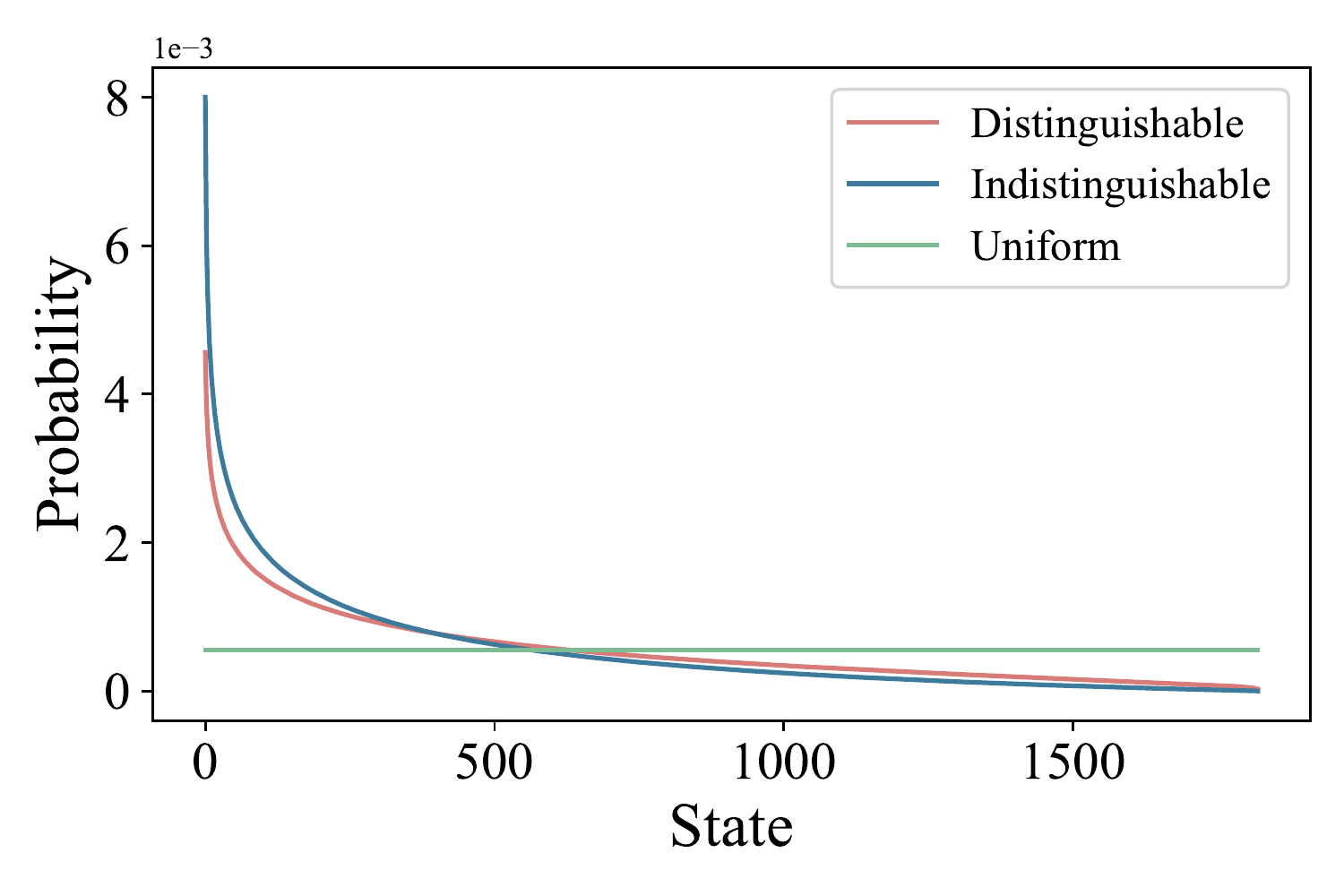}
	\caption{\label{fig:profiles} Average sorted profiles for distinguishable and indistinguishable regimes of boson sampler with 4 photons and 16 output modes. Probability corresponding to uniform distribution is denoted as green line for reference.}
\end{figure}

In this work we used \texttt{Boson-Sampling} python package~\cite{samplersoft} to calculate probability distributions for a given boson sampler matrix $U$ that is a $m\times n$ matrix that describe a relation between creation operators of input ($\hat{a}^{\dagger}_j$) and output ($\hat{b}^{\dagger}_i$) modes. The Haar distributed complex unitary interferometer matrix itself was generated by using \texttt{Strawberry Fields} python package~\cite{Bromley:2020aa,Killoran2019strawberryfields} which utilizes classical groups approach~\cite{unitarygeneration}. 
In Fig.~\ref{fig:profiles} we compare the sorted probability distributions for collision-free outcomes obtained from distinguishable, indistinguishable and uniform boson samplers. One can see that there is a little difference between the profiles obtained with distinguishable and indistinguishable particles, which clearly demonstrates the complexity of the source benchmarking with a limited number of outcomes. In general, to reproduce these probability profiles by using the statistics of the outcomes, the number of bitstrings should be four or five times larger than the size of the state space.  

\begin{figure}[t]
    \includegraphics[width=\columnwidth]{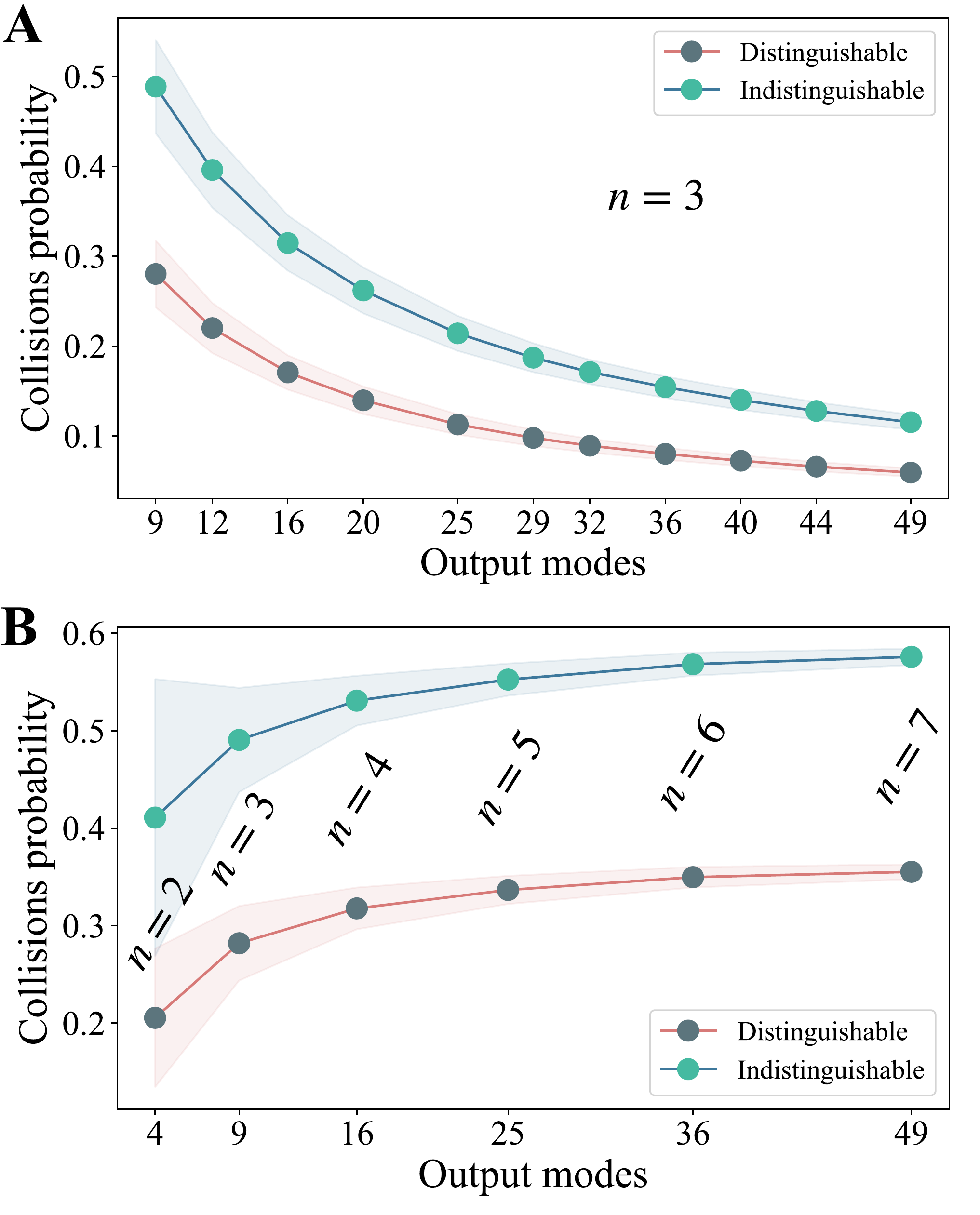}
	\caption{\label{collisions} Overall probability of generating outcome with collisions. (A) Results obtained with fixed number of bosons, $n = 3$ and different numbers of output modes. (B) The number of bosons and output modes are varied as  $n = \sqrt{m}$, where $m$ in the number of output modes. The data for each setting were averaged over 1000 scattering matrices. Colored regions denote the corresponding standard deviations.}
\end{figure}

Detecting the collision events in the real experiments represents a significant technological challenge, which makes selecting the collision-free outcomes a natural solution when benchmarking BS device. Since neglecting collision events leads to considerable changes in the behaviour of the correlation-based features as it was demonstrated in Figs.~\ref{fig:clouds} C and D in the main text, it is important to estimate the probabilities of generating an event with collisions in different BS settings. For instance, in Fig.~\ref{collisions} A we show such a collision outcome probability as a function of the number of output modes at the fixed number of bosons. As one would expect the probability decays fast as $m$ increases.

The situation is completely different if number of output modes depends on the number of bosons as $m = n^2$ (Fig.~\ref{collisions} B), which can be associated to the lower bound for demonstrating quantum advantage. In this case the collision probability for indistinguishable particles does not decreases with $m$. Instead it is saturated at a rather high value of about 0.55, which suggests a careful examination of the results of the previous correlation-function-based studies \cite{Walschaers2016, Flamini_2020} when applying them for analysis of real experiments. Another important finding is that the collision probability for distinguishable particles is about two times smaller than that for indistinguishable ones, which might be also used for certifying a BS device. 

\begin{table}[b]
    \centering
    \caption{Percentage of repetitions $(N-N_{meas})/N_{meas}~\,(\%)$ in the data set with N unique bitstrings in both BS regimes. The results are averaged over 100 random $U$ matrices.}\label{tab:repetitions}
    \begin{tabular}{cccc}
    \hline
    \hline
    $N$\,& $n$\, & $\mspace{8mu}$distinguishable$\mspace{8mu}$ & $\mspace{8mu}$indistinguishable$\mspace{8mu}$ \\\hline
      512 & 4 & 27.715 & 37.419 \\ 
      \, & 5 & 1.019 & 1.605  \\
      \, & 6 & 0.018 & 0.057 \\ 
      \, & 7 & 0.0 & 0.002 \\
      1024 & 4 & 54.045 & 67.680 \\
      \, & 5 & 1.957 & 3.322  \\ 
      \, & 6 & 0.055 & 0.095 \\
      \, & 7 & 0.002 & 0.001 \\\hline \hline
    \end{tabular}
\end{table}

In contrast to the collisions, the amount of repetitions drastically decreases with $n$. As can be seen from Tab.~\ref{tab:repetitions}, bitstrings collected from BS with $n\ge 6$ are almost unique up to some moderate number $N$. This is obviously connected with the rapid grows of the state space. Thus, excluding repetitions plays role only when considering $n=4$ sampler.  

\subsection*{Choice of the cutoff radii}
In this section we discuss the choice of the cutoff radius when constructing Hamming nets. As it was shown in Ref.~\onlinecite{Heyl2023} one should avoid overcounting isolated nodes and keeps a non-trivial network structure. Undoubtedly, such a choice is problem specific and in our study we are looking for the $R$ values at which properties of the distinguishable network differs as much as possible from that obtained with indistinguishable bosons. We calculated the degree distribution that defines a fraction of nodes with $k$ connections to other nodes and is standard measure to characterize a graph or network. We find out that the most reliable features are two first moments of the $P_{k}$ distributions for such intermediate cutoff radii for which $P_{k}$ is Gaussian-like and has both left and right sides. 

\begin{figure*}[t]
    \includegraphics[width=0.9\textwidth]{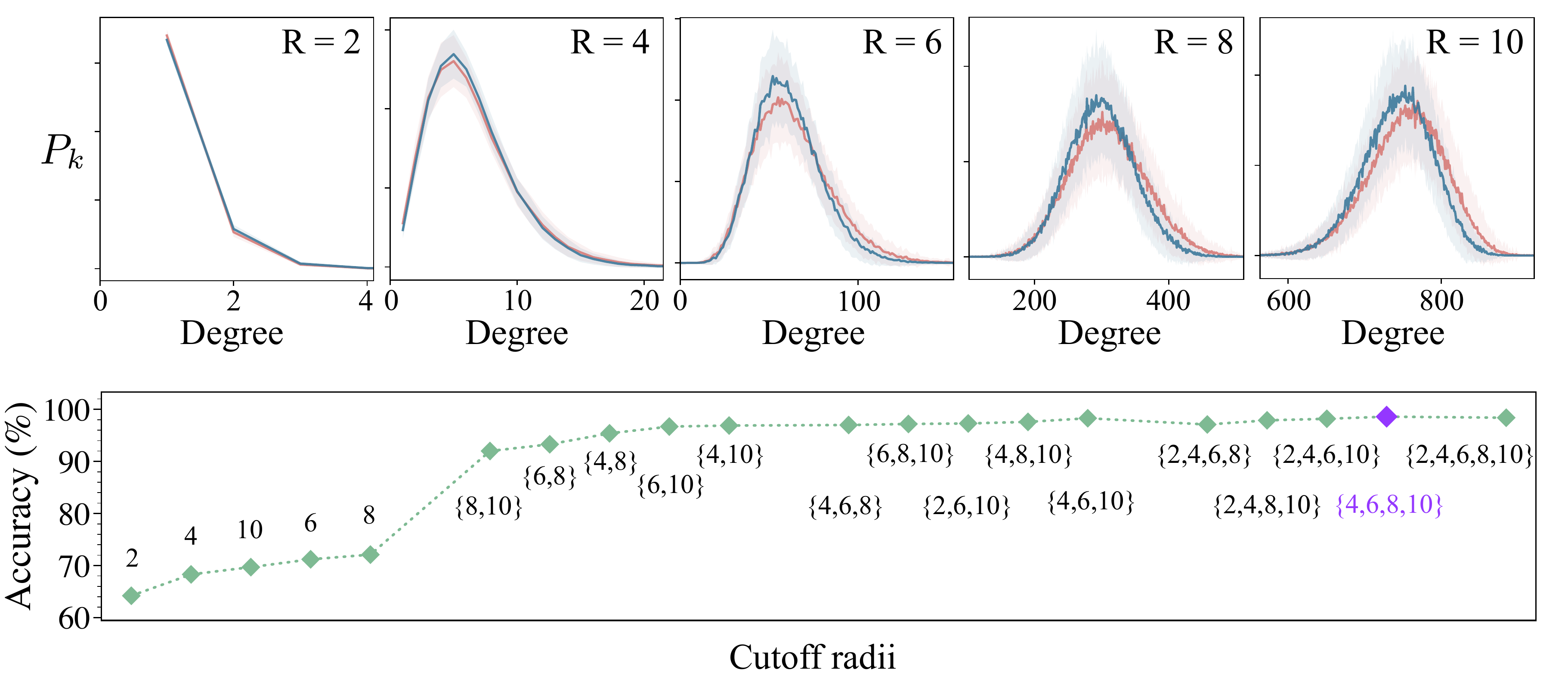}
	\caption{\label{fig:acc_vs_R} (Top) Comparison of the averaged distributions for indistinguishable (blue) and distinguishable (red) bosons, obtained at different cutoff radii using $N=1024$ unique bitstrings from a boson sampler with $n$ = 6 and $m$ =36. All the results are averaged over 100 random $U$ matrices. (Bottom) Accuracy dependence on the cutoff radii used to define feature vector. The best accuracy is marked by purple diamond and equal to 98.6\%. Here we used up to five combinations in each number of cutoff radii in each group based on the obtained precision.}
\end{figure*}

Importantly, taking into account distributions at small $R$ may also be useful since mean values of $P_{k}$ distributions are rather robust and then it gives some new dimensions to the feature space where there is at least some discrepancy between the different BS regimes. However, as can be seen from Fig.~\ref{fig:acc_vs_R}, this is not the case if we already took into account all the meaningful cutoff radii.
Thus, we use $R=\{2, 4, 6\}$, $R=\{2, 4, 6, 8\}$, $R=\{4, 6, 8, 10\}$ and $R=\{6, 8, 10, 12\}$ for the samplers with $n$ = 4, 5, 6 and 7 photons, respectively. 

\begin{figure}[t]
    \includegraphics[width=0.9\columnwidth]{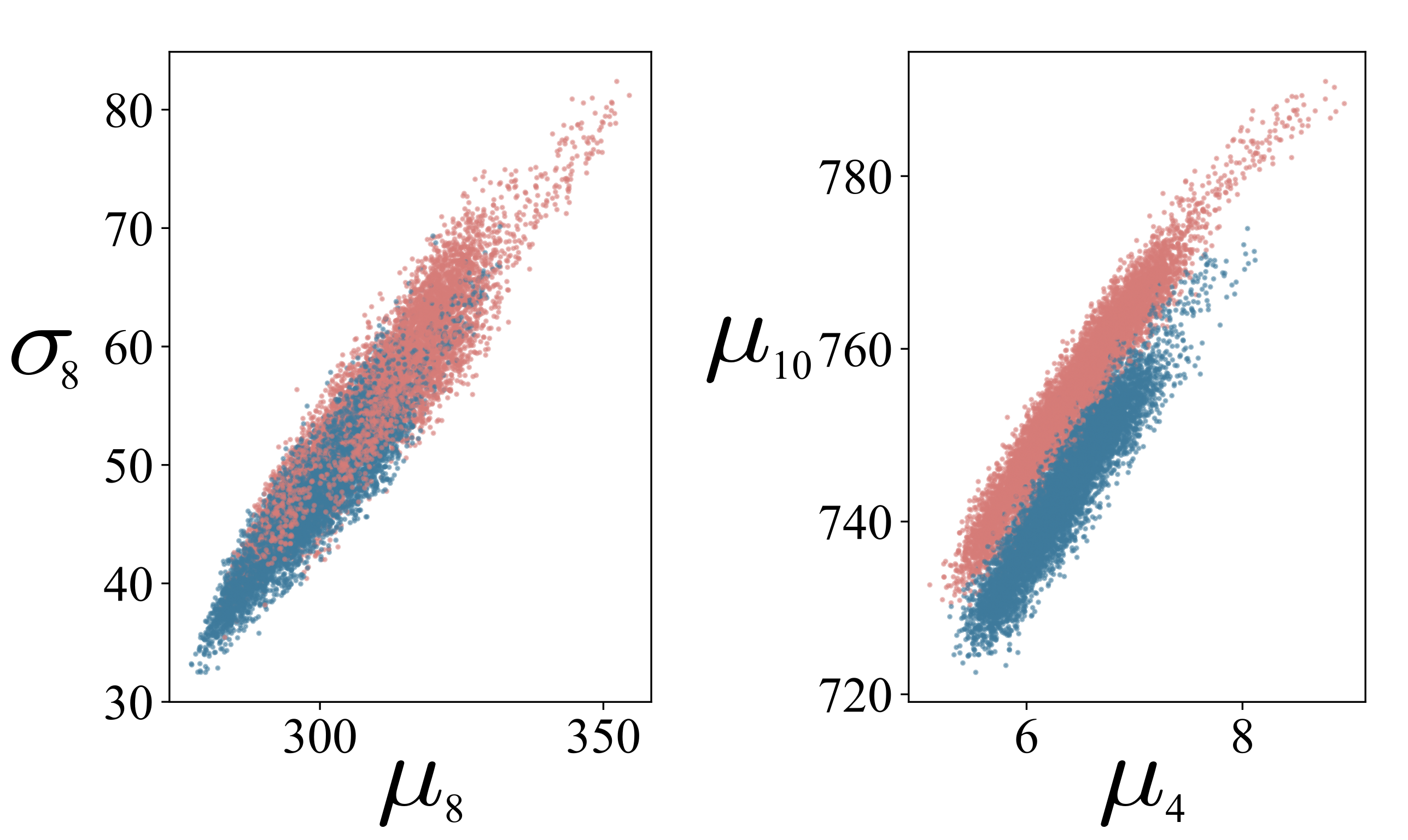}
	\caption{\label{fig:s_clouds} Low-dimensional visualisations of the configurations taken from the training sets for $n=6$ distinguishable (red) and indistinguishable (blue) bosons and $m=36$ output modes, $N=1024$. Features were taken from the Hamming networks providing the best possible accuracy on a single $R$ (left) and pair of cutoff radii (right). Namely, $R=8$ and $R=\{4, 10\}$.}
\end{figure}

The reason why the accuracy is low when taking into account single cutoff radius is that the clouds formed by different samplers strongly overlap in $\sigma$-$\mu$ plane. As can be seen from Fig.~\ref{fig:s_clouds}, adding at least one extra $R$ leads to the formation of almost separated clouds in $\mu_1$-$\mu_2$ plane. We should stress that both moments of $P_{k}$ distributions are essential since it leads to better separation of the configurations in the higher-dimensional parameter space.    

\begin{table}[t]
    \centering
    \caption{Comparison of the accuracy (\%) of different ML classifiers trained with the same data set on the unseen $U$ matrices. Features were taken from both Hamming network and mode-mode correlations.}\label{tab:ml_comparison}
    \begin{tabular}{cccccc}
    \hline
    \hline
    $N$\,& $n$\, & $\mspace{8mu}$LR$\mspace{8mu}$ & $\mspace{8mu}$SVM$\mspace{8mu}$ & $\mspace{8mu}$RF$\mspace{8mu}$ & $\mspace{8mu}$$k$-NN$\mspace{8mu}$ \\\hline
      512 & 4 & 95.6(1) & 94.3(1) & 93.3(1) & 93.0(1) \\ 
      \, & 5 & 97.1(1) & 96.9(1) & 96.4(1) & 96.6(1) \\
      \, & 6 & 91.8(1) & 91.6(1) & 90.9(1) & 90.7(1) \\ 
      \, & 7 & 89.0(1) & 88.9(1) & 88.6(1) & 88.7(1)\\
      1024 & 4 & 100.0 & 100.0 & 100.0 & 100.0\\
      \, & 5 & 99.8(1) & 99.8(1) & 99.7(1) & 99.6(1) \\ 
      \, & 6 & 98.9(1) & 98.8(1) & 98.5(1) & 98.6(1)\\
      \, & 7 & 98.3(1) & 98.2(1) & 98.2(1) & 98.1(1)\\\hline \hline
    \end{tabular}
\end{table}

\subsection*{Machine learning }

To solve the problem of distinguishing between two different BS regimes we train several basic ML algorithms implemented in \texttt{Scikit-learn} python package~\cite{scikit-learn}.
Namely, a logistic regression (LR),  a support vector machine (SVM), random forest (RF) and $k$-nearest neighbours ($k$-NN) classifiers. For LR we use the \textit{liblinear} solver, for SVM -- radial basis function as a kernel and $\gamma=1/N_f$, where $N_f$ its the length of our feature vector, and for RF -- 300 estimators. The amount of nearest neighbours in $k$-NN we adjust manually for each data set to get the best accuracy. The rest parameters of the algorithms were chosen to be the default ones. 

The main data set includes 100 samples for each regime for each of 80 different $U$ matrices. We randomly shuffle this data and use 80\% as a training set and 20\% as a testing one. To check the accuracy on the unseen data we generate and use additional 4000 samples from 20 completely new random U matrices. Since each feature lies in its own range and has its own dispersion, additional standardization is done on the basis of the training samples.  

As can be seen from Table~\ref{tab:ml_comparison}, all the presented algorithms show similar performance on the unseen data when we use features taken from both Hamming network and mode-mode correlations. However, LR is more stable, faster and slightly outperforms the rest ones in most cases which makes it more preferable for the analysis of currently available interferometers.


\bibliography{bibliography.bib}

\end{document}